\documentclass[a4paper,10pt,twoside]{cpc-hepnp}

\usepackage[hang,tight]{subfigure}
\usepackage{threeparttable}
\usepackage{multicol}
\usepackage{graphicx}
\usepackage{booktabs}
\usepackage{amssymb,bm,mathrsfs,bbm,amscd}
\usepackage[tbtags]{amsmath}
\usepackage{lastpage}

\begin{document}

\fancyhead[c]{\small Chinese Physics C~~~Vol. 37, No. 1 (2013) 010201}

\footnotetext[0]{*Supported by National Natural Science Foundation of China (11105160)
}

\title{Neutron collimator design of neutron radiography based on the BNCT facility}

\author{%
      YANG Xiao-Peng(ÑîÏþÅô)$^{1;1)}$\email{yangxiaopeng@ihep.ac.cn}%
      \quad YU Bo-Xiang(Óá²®Ïé)$^{2;2)}$\email{yubx@ihep.ac.cn(corresponding author)}%
       \quad Li Yi-Guo(ÀîÒå¹ú)$^{3}$%
       \quad Peng Dan(Åíµ©)$^{3}$\\%
       \quad Lu Jin(³½÷)$^{3}$%
       \quad Zhang Gao-Long(ÕŸßÁú)$^{1}$%
       \quad ZHAO Hang(ÕÔº½)$^{2}$%
       \quad ZHANG Ai-Wu(Õ°®Îä)$^{2}$\\%
       \quad LI Chun-Yang(Àî´ºÑô)$^{4}$%
      \quad LIU Wan-Jin(ÁõÍò½ð)$^{2}$%
      \quad HU Tao(ºúÌÎ)$^{2}$%
      \quad L\"U Jun-Guang(ÂÀ¾ü¹â)$^{2}$%
}
\maketitle

\address{%
1~(School of Physics and Nuclear Energy Engineering, Beihang University, Beijing 100191,China)\\
2~(State Key Laboratory of Particle Detection and Electronics, Institute of High Energy Physics,CAS)£¬ Beijing 100049, China)\\
3~(China Institute of Atomic Energy, Beijing 102413, China)\\
4~(University of South China, Hengyang 421000, China£©\\
}

\begin{abstract}
For the research of CCD neutron radiography, a neutron collimator was designed based on the exit of thermal neutron of the Boron Neutron Capture Therapy (BNCT) reactor. Based on the Geant4 simulations, the preliminary choice of the size of the collimator was determined. The materials were selected according to the literature data. Then, a collimator was constructed and tested on site. The results of experiment and simulation show that the thermal neutron flux at the end of the neutron collimator is greater than $1.0\times10^6$ n/cm$^{2}$/s, the maximum collimation ratio~(L/D) is 58, the Cd-ratio(Mn) is 160 and the diameter of collimator end is 10 cm. This neutron collimator is considered to be applicable for neutron radiography.

\end{abstract}



\begin{multicols}{2}
\section{Introduction}
X ray radiography is a powerful non-destructive medical tool to view a non-uniformly composed material such as the human body and obtain information on the distribution of minerals, fractures, stratification, and other structural features. This technology is also applied for neutron radiography~\cite{lab1,lab2} due to some of the same characteristics for thermal neutrons and X ray such as no charge and a similar wavelength with object atoms. Radiography technology is mainly based on ray attenuation when the rays go through the object. When X ray goes through the object, it interacts with electrons. So the absorption ability of the object for X ray is related to the number of extranuclear electron. However, the neutron attenuation is different from that of X ray. It can be attributed to the scattering between neutrons and object nuclei or be directly absorbed by the object nuclei. Thus the neutron attenuation has nothing to do with the atomic number of the object. The X ray radiography cannot separate two different kinds of objects with the close atomic number, but neutron radiography can obviously distinguish them due to the different mechanisms with these objects. Hence neutron and X ray radiographies can compensate each other to distinguish the different irradiated objects. Moreover the tomography can be applied based on the radiography operating a three-dimensional reconstruction from the two-dimensional transmission images allowed from different view angles by the radiography.

Like X ray radiography, the suitable radioactive source should be provided for neutron radiography. To obtain fast radiographic images and good position resolution of neutron radiography, the necessary neutron source used today should be provided by nuclear reactor or spallation neutron sources followed by a neutron collimator~\cite{lab2,lab3,lab4}. Furthermore ideal neutron beams for neutron radiography should be parallel, monoenergetic, high flux possessing no other radiation with good uniformity at its cross section. However, actually, what we can do is to try our best to be more close to these ideal parameters. So an ideal thermal neutron collimator can produce good quality neutron beam with high collimation ratio~(higher than 100), low gamma background (as low as possible), high Cd-ratio(Mn)(higher than 200) and well-distributed neutron flux (like square wave) at the exit of the collimator.

This work is to design an appropriate neutron collimator at the thermal neutron exit of the BNCT reactor facility for the research of neutron radiography based on a CCD camera.

\section{Geant4 simulation for the thermal neutron collimator}

Neutron radiography has been studied by many researchers and many different neutron radiography facilities have been developed~\cite{lab2,lab3,lab4}. According to the present research and the structure of the BNCT reactor facility, the following sections will describe the simulation of the neutron collimator.

The design of the neutron radiography collimator is shown in Fig.1. The left part is the design drawing of the intrinsic thermal neutron exit of the BNCT reactor facility and the right part is the collimator designed particularly for neutron radiography. The distance from the reactor core to the thermal neutron exit of the BNCT reactor facility is 1.615 m. The total length of the collimator is 1.15 m consisting of one bismuth gamma filter, five boron rings and a series of lead parts. During the design process of the neutron collimator, the Monte Carlo method is used to simulate the whole thermal neutron collimator. By the simulation, the better size of units is determined.
\begin{center}
\includegraphics[width=8cm,height=4.8cm]{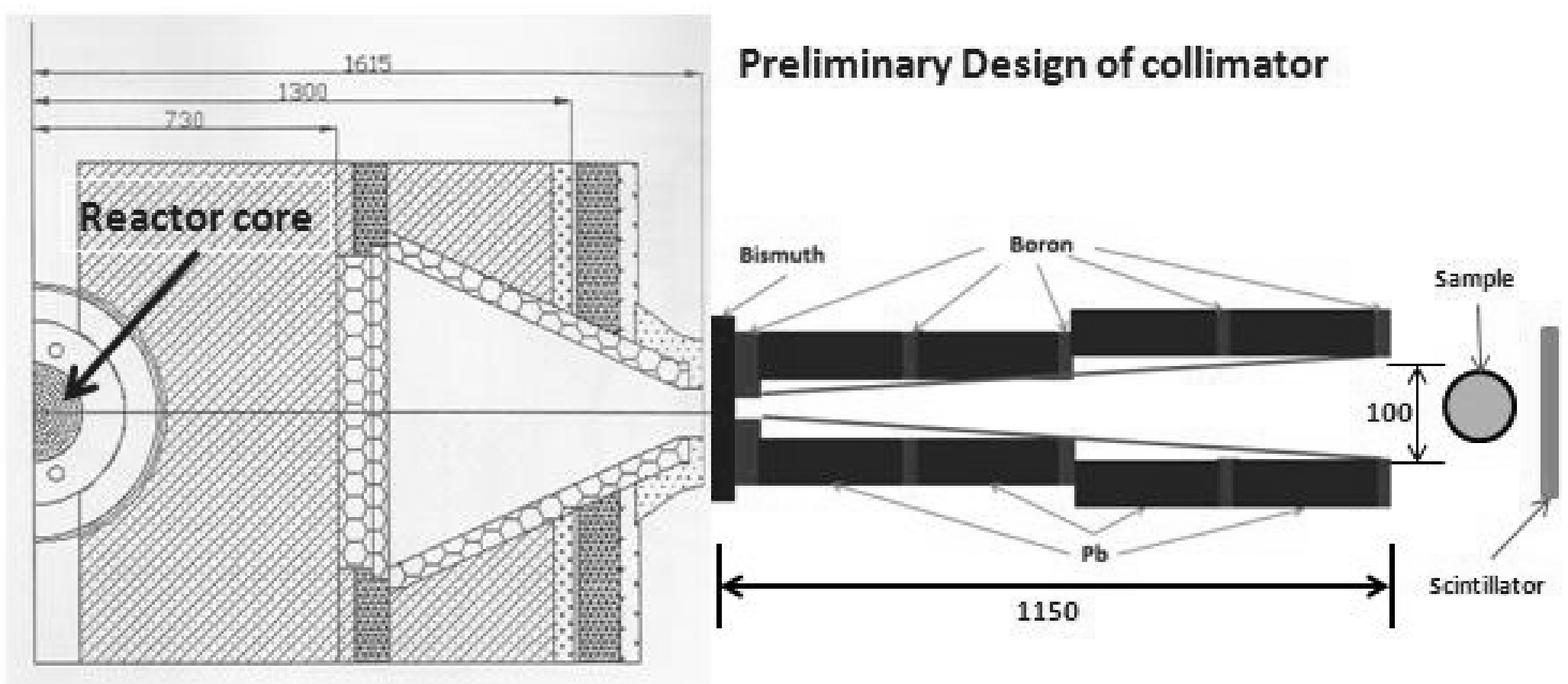}
\figcaption{\label{fig1} Design drawing of the thermal radiography collimator.}
\end{center}

\subsection{Introduction of the physical model needed for simulation}

The strong interaction model QGSP$\_$BERT$\_$HP includes the cross section library of the thermal neutron, so it is selected in the simulation of the radiography collimator. Furthermore, it shows an excellent agreement with the MCNP results. Hence this model is recommended for the application of the industry, medicine, radio shielding and dosimetry by the project cooperation group of GEANT4. Moreover, there are many successful applications~\cite{lab2,lab3}.

\subsection{Design requirements for the radiography collimator and simulation contents}

According to References \cite{lab2,lab3,lab4,lab5} of the radiography collimator and the current situation of the thermal neutron exit of the BNCT reactor facility, a series of design requirements for the thermal radiography collimator are proposed. The requirements are as follows:

\noindent- Thermal neutron flux at the exit of the collimator should be more than $5\times10^5$n/cm$^{2}$/s;

\noindent- The collimation ratio (L/D) is limited to two options according to the factual situation;

\noindent- The diameter of the collimator exit is 10 cm;

\noindent- The ratio of neutrons to gamma rays is more than $10^5$ n/cm$^{2}$/mrem;

\noindent- The cadmium ratio is more than 150.

During the simulation of the source input, the GPS module is used to describe the thermal neutron at the exit of the BNCT reactor facility as the neutron source. Meanwhile the energy spectra and angular distributions of neutrons and gamma-rays at the thermal neutron exit of the BNCT reactor facility for the original reactor simulation are adopted. According to measurement, the flux distribution of the thermal neutron at the exit of the BNCT reactor facility is basically consistent with the numerical simulation, which reaches up to $1.9\times10^9$ n/(cm$^{2}$$s^{-1}$) and has good uniformity in the range of 12 cm in diameter.
\begin{center}
\includegraphics[width=8cm,height=4.8cm]{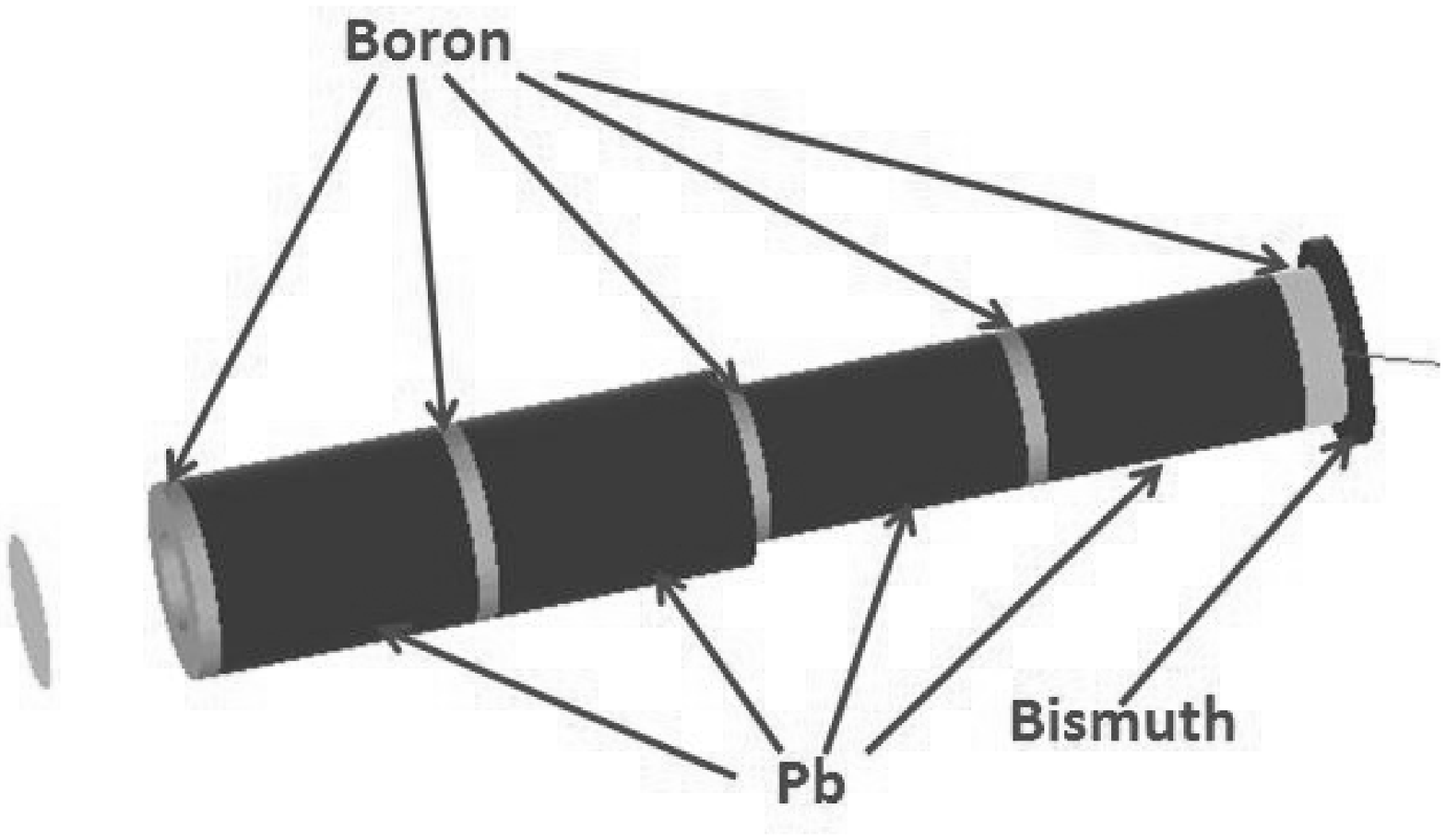}
\figcaption{\label{fig2} The structure of neutron collimator based on GEANT4 simulation.}
\end{center}

Additional constraints are needed in the design of the radiography collimator to match the designed parameters. The bismuth block with certain thickness is placed at the thermal neutron exit of the BNCT reactor facility in order to minimize the gamma rays background and scatter off some fast neutrons. Following that a conic-shaped collimator (Fig.2) comprising the apertures with continuous interactive boron rings and lead parts is adopted to collimate the thermal neutron from the source towards the exit of the collimator and reduce the gamma ray background. Due to the larger cadmium ratio (Cd-ratio(Mn)=160.05) at the thermal neutron exit of the BNCT reactor facility, a fast neutron filter isn't needed. The aim of simulation is to optimize the size and the position of the collimator parts to get the good thermal neutron quality.

\subsection{The simulation results and discussions}

The major parameters of the thermal neutron collimator are the collimation ratio (L/D), the Cd-ratio, the neutron flux at the collimator exit and the gamma dose rate. To obtain the appropriate L/D, the diameter of the aperture is proposed to 2 cm or 4 cm. Because the aperture made of boron is nearly transparent to gamma rays, the aperture size does't affect the gamma dose rate (Fig.3.b). Fig.3 shows the change of neutron flux and gamma dose rate with the thickness of bismuth. From Fig. 3, when the bismuth thickness increases, the neutron flux and gamma dose rate show an exponential decline. In order to match the requirements: the neutron flux reaching $5\times10^5$ n/cm$^{2}$/s and the lowest corresponding gamma dose rate, simulation results show that the optimal bismuth thickness is 3 cm. Under this condition, the gamma dose rate is 2.1 mrem/s and the ratios of the neutron to gamma with the aperture diameter of 2 cm and 4 cm are $3.0\times10^5$ n/cm$^{2}$/mrem and $6.1\times10^5$ n/cm$^{2}$/mrem respectively.

\end{multicols}

\begin{figure}[!htb]
  \begin{center}
  \vspace{-0.5cm}
        \subfigure[]{
          \label{fig3:a}
          \begin{minipage}[b]{0.5\textwidth}
            \centering
            \includegraphics[width=7.5cm,height=4.7cm]{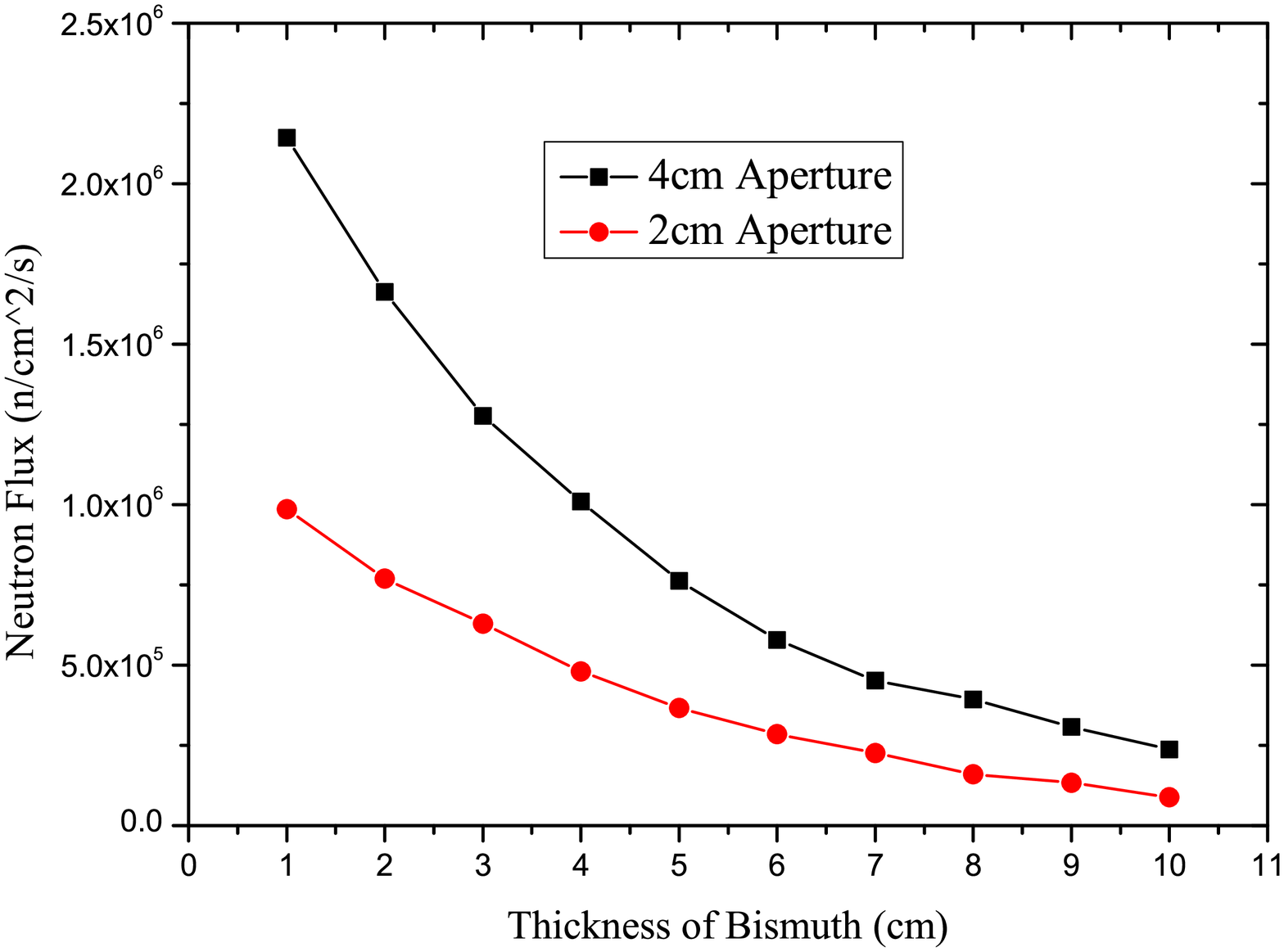}
          \end{minipage}}%
        \subfigure[]{
          \label{fig3:b}
          \begin{minipage}[b]{0.5\textwidth}
            \centering
            \includegraphics[width=7.5cm,height=4.7cm]{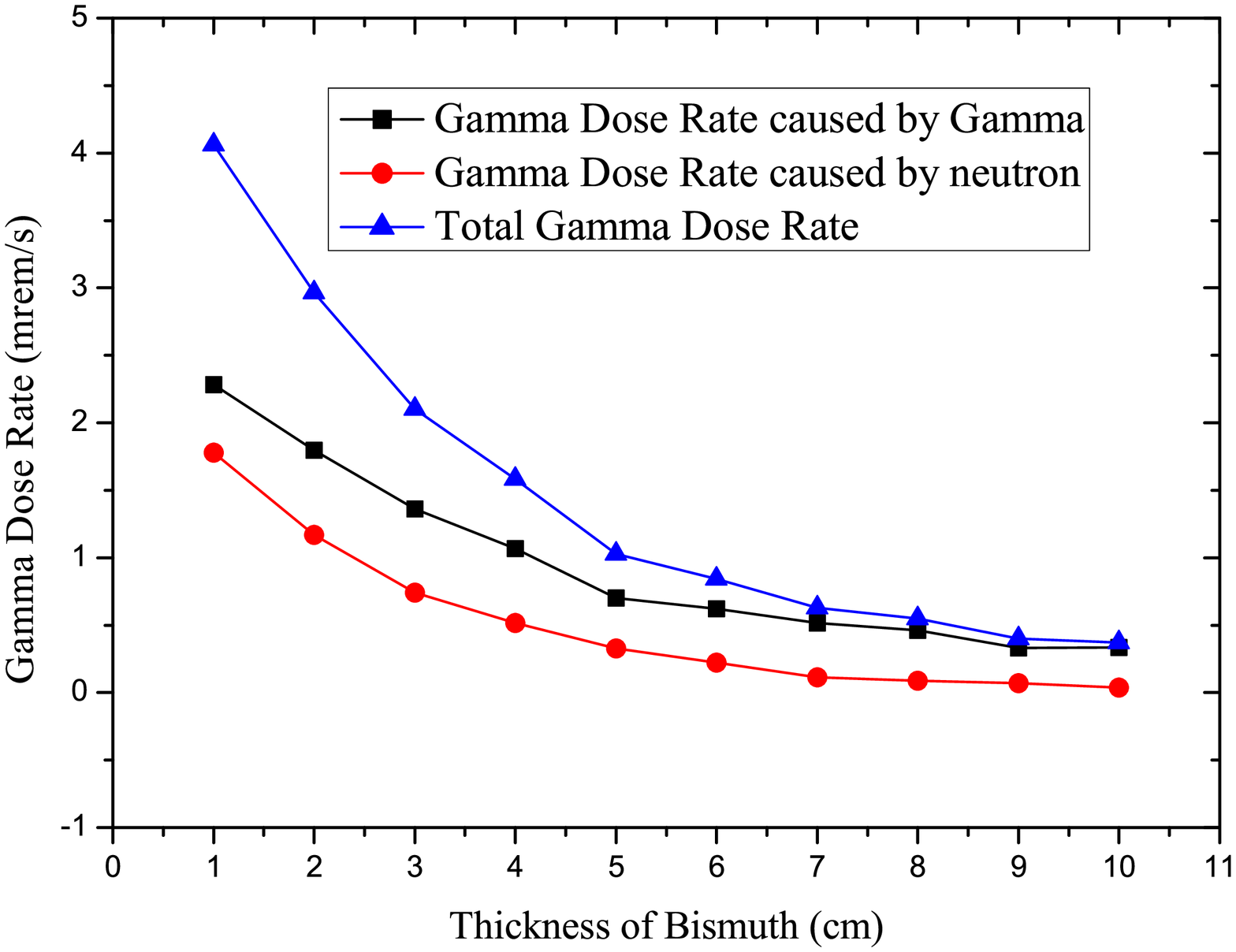}
          \end{minipage}}%
        \caption{The neutron flux and gamma dose rate at the exit of the collimator varying with the increase of bismuth thickness. (a)~The changes of the neutron flux follow with the change of bismuth thickness. \quad(b)~The changes of the gamma dose rate follow with the change of bismuth thickness.}
        \label{fig3}
  \end{center}
\end{figure}
\vspace{-1.3cm}
\begin{multicols}{2}

For 2 cm and 4 cm aperture the thermal neutron flux distributions by the normalization of the neutron flux at the center at the exit of the collimator have little difference. The thermal neutron flux distribution is shown in Fig. 4. We can see that for the original source the flux uniformity is poor at the collimator exit, however, within the range of 2 cm radius the uniformity is better. At the 4 cm radius the neutron flux is about 30\% of the neutron flux at the center. When we change the incoming neutron angular distribution, the neutron flux uniformity is greatly improved. It is found that the angular distribution of  incident neutron will influence the flux uniformity significantly.
\begin{center}
\includegraphics[width=8cm,height=4.7cm]{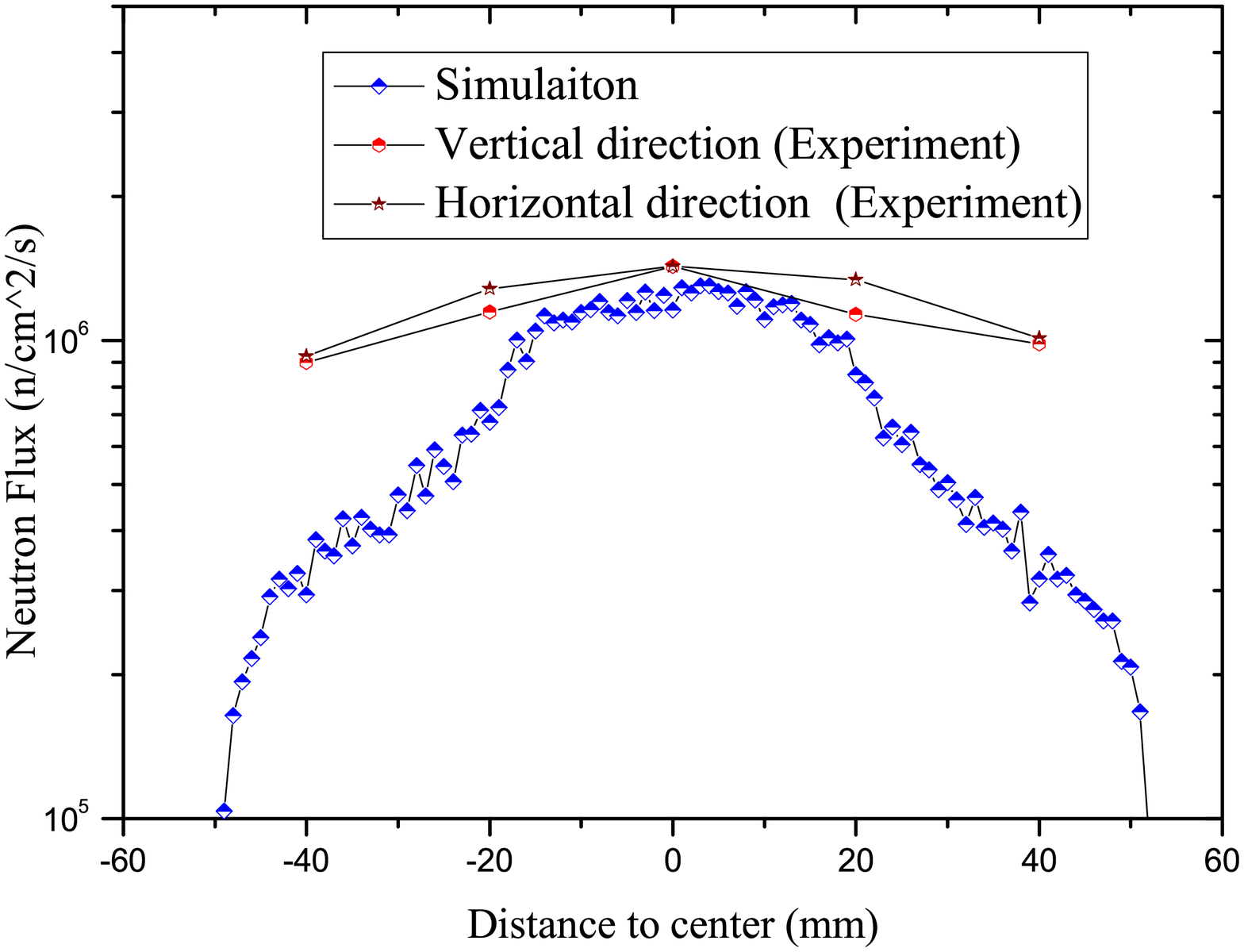}
\figcaption{\label{fig4} The thermal neutron flux distribution of the radiography collimator based on the GEANT4 simulation and the measurement result for 4 cm aperture.}
\end{center}

Through the detailed simulation and analysis the expected characteristics of the neutron radiography collimator are presented in Table 1.

\begin{center}
\tabcaption{ \label{tab1} Simulation results of the collimator.}
\footnotesize
\begin{tabular*}{80mm}{c@{\extracolsep{\fill}}cc}
\toprule Parameter name & Value\\
\hline
Total length (mm): & 1150\\
Collimation ratio (L/D): & 29 or 58\\
Apertures (cm): & 4 or 2\\
Cd ratio (Mn): & 160\\
Neutron flux at the collimator exit(n/cm$^{2}$/s): & $1.3\times10^6$or$ 6.3\times10^5$\\
Bismuth filter thickness (cm): & 3\\
Beam diameter (cm): & 10\\
Divergence angle (deg.): & 4.1\\
Gamma dose rate (mrem/s): & 2.1\\
\bottomrule
\end{tabular*}
\vspace{0mm}
\end{center}
\vspace{0mm}

\section{Measurement of the thermal neutron collimator }

After finishing the GEANT4 simulation of the neutron radiography collimator, a series of components are produced according to the design. The collimator is composed of the lead brick, boron rings and some other units as shown in Fig 5. At the thermal neutron exit of the BNCT reactor facility, 3-cm thick bismuth is selected~\cite{lab5,lab6}. Next to the bismuth filter is the boron aperture with the diameter of 4 cm or 2 cm to restrain the thermal neutron. Behind the bismuth filter a series of boron rings and lead parts are available for the neutron shaping and gamma shielding.
\begin{center}
\includegraphics[width=8cm,height=5cm]{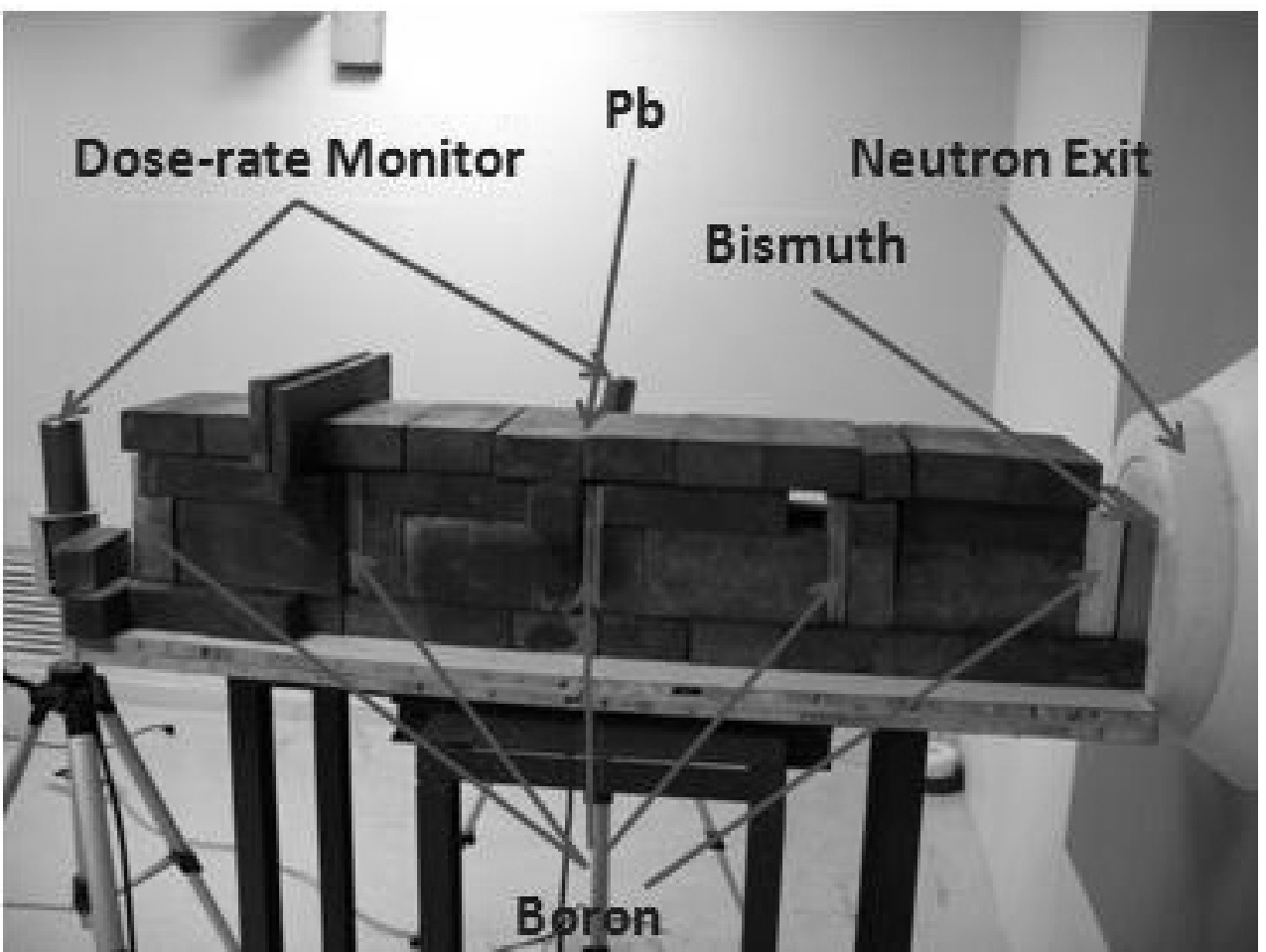}
\figcaption{\label{fig5} Test of the radiography collimator.}
\end{center}

During the experiment, the neutron detectors formed by the manganese activation pieces were located at the exit of the radiography collimator to obtain the neutron flux distribution. The gamma dose rate at the exit of the collimator was detected by the thermoluminescent dose detectors.

The distribution of neutron flux at the collimator exit was measured for the 4 cm diameter aperture, which is shown in Fig 4. It shows that within the range of 2 cm radius the neutron flux remains stable, the neutron flux at the radius of 4 cm is about 60\% of that at the radius of 2 cm and the maximum neutron flux at the collimator center is $1.43\times10^6$ n/cm$^{2}$/s. If the collimator entrance is replaced by the boron aperture with the 2 cm diameter, the distribution of thermal neutron flux doesn't change significantly. But the neutron flux at the collimator center is only $5.4\times10^5$ n/cm$^{2}$/s. The experimental result shows a reasonable agreement with the simulation result within the 4 cm diameter but bad consistency out of the 4 cm diameter which can be influenced by the angular distribution of incident neutron.

At the same time the gamma dose rates at the collimator center and around the collimator are tested for the 4 cm diameter aperture. The gamma dose rate at the collimator center is little higher due to insufficient lead shielding, which can reach 3.6 mrem/s bigger than the simulation result 2.1mrem/s. So the ratio of neutrons to gamma rays is about $4\times10^5$ n/cm$^{2}$/mrem higher than the requirement ratio $10^5$ n/cm$^{2}$/mrem. Meanwhile, the Cd-ratio(Mn) tested is 160 better than the original value 150. Hence in the future experiment more lead shields inside the collimator are needed to reduce the original gamma-ray background and the gamma rays produced by the interaction between the thermal neutron and the boron.

\section{ Conclusion}

Through the GEANT4 simulation of the thermal neutron radiography collimator the important parameters are optimized and finally determined such as the thermal neutron flux, the cadmium ratio, the collimation ratio, the gamma dose rate, the size of the collimator parts, etc. According to the parameters obtained by the simulation, the parts are produced and the neutron radiography collimator is constructed at the thermal neutron exit of the BNCT reactor facility. Then the neutron flux and the gamma dose rate are measured when two different apertures are employed. The measured flux of the thermal neutron basically agrees to the simulation results. Hence, the collimator can be applied in neutron radiography.

\end{multicols}

\vspace{-1mm}
\centerline{\rule{80mm}{0.1pt}}
\vspace{2mm}

\begin{multicols}{2}

\end{multicols}

\clearpage


\begin{thebibliography}{90}

\vspace{3mm}

\bibitem{lab1}
Pleinert H, Lehmann E, K\"orner S. Nucl. Instr. And Meth. A, 1997, 399: 2-3
\bibitem{lab2}
Schillinger B, Calzada E, Gr\"unauer F et al. Applied Radiation Isotopes, 2004, 61: 4
\bibitem{lab3}
Burgio N, Rosaet R. Applied Radiation Isotopes, 2004, 61: 4
\bibitem{lab4}
Ouardi A, Machmach A, Alami R et al. Nucl. Instr. And Meth. A,2011£¬651£º1
\bibitem{lab5}
M. DINCA, PAVELESCU M, IORGULIS C. Rom. Journ. Phys., 2006, 51: 3-4
\bibitem{lab6}
JIANG Xin-Biao, ZHU Yang-Ni. The Report Of The Neutron Irradiator In The Hospital, 2010
\end{thebibliography}
\end{document}